%% file: wsc23paper.tex
\documentclass{wscpaperproc}
\usepackage{latexsym}
\usepackage{graphicx}
\usepackage{mathptmx}
\usepackage[T1]{fontenc}

\usepackage{amsmath}
\usepackage{amsfonts}
\usepackage{amssymb}
\usepackage{amsbsy}
\usepackage{amsthm}
\usepackage{comment}
\usepackage{subfig}
\usepackage{interval}

\usepackage[pdftex,colorlinks=true,urlcolor=blue,citecolor=black,anchorcolor=black,linkcolor=black]{hyperref}

\usepackage[capitalise]{cleveref}

\newtheoremstyle{wsc}
{3pt}
{3pt}
{}
{}
{\bf}
{}
{.5em}
{}

\theoremstyle{wsc}

\begin{document}


\pagestyle{fancyplain}

\thispagestyle{plain}
\firstPageHead{}

\chead{\fancyplain{}{\itshape Ahmed, Rahimian, and Roberts}}

\rhead{}
\cfoot{}
\renewcommand{\headrulewidth}{0pt} 

\input{wscbib.tex}           

\setlength{\baselineskip}{12.7pt}

\title{INFERRING EPIDEMIC DYNAMICS USING GAUSSIAN PROCESS EMULATION OF AGENT-BASED SIMULATIONS}

\author{Abdulrahman A. Ahmed\\ [12pt]
	Department of Industrial Engineering\\
	University of Pittsburgh\\
	3700 O'Hara Street\\
	Pittsburgh, PA 15261, USA\\
\and 
M. Amin Rahimian\\ [12pt]
Department of Industrial Engineering\\
University of Pittsburgh\\
3700 O'Hara Street\\
Pittsburgh, PA 15261, USA\\
\and
Mark S. Roberts\\ [12pt]
Department of Health Policy and Management\\
University of Pittsburgh\\
130 Desoto Street\\
Pittsburgh, PA 15261, USA\\
}

\maketitle

\section*{ABSTRACT}

Computational models help decision makers understand epidemic dynamics to optimize  public health interventions. Agent-based simulation of disease spread in synthetic populations allows us  to compare and contrast different effects across identical populations or to investigate the effect of interventions keeping every other factor constant between "digital twins". In particular, we can observe the behavior of two different diseases as they evolve from identical initial conditions in the same population. FRED (A Framework for Reconstructing Epidemiological Dynamics) is an agent-based modeling system with a geo-spatial perspective using a synthetic population that is constructed based on the U.S. census data. Having synthetic data provides a baseline to get comparable results from different conditions and interventions. In this paper we show how Gaussian process regression can be used on FRED-synthesized data to infer the differing spatial dispersion of the epidemic dynamics for two disease conditions that start from the same initial conditions and spread among identical populations. Our results showcase the utility of agent-based simulations frameworks such as FRED for inferring differences between conditions where controlling for all confounding factors for such comparisons is next to impossible without synthetic data. We also show the advantage of surrogate models such as Gaussian process regression when the computational cost of inference with synthetic data alone is foreboding.

\input{body}

\end{document}

%% file: wscbib.tex
\makeatletter
\let\@internalcite\cite
\def\cite{\def\@citeseppen{-1000}%
    \def\@cite##1##2{(##1\if@tempswa , ##2\fi)}%
    \def\citeauthoryear##1##2##3{##1 ##3}\@internalcite}
\def\citeNP{\def\@citeseppen{-1000}%
    \def\@cite##1##2{##1\if@tempswa , ##2\fi}%
    \def\citeauthoryear##1##2##3{##1 ##3}\@internalcite}
\def\citeN{\def\@citeseppen{-1000}%
    \def\@cite##1##2{##1\if@tempswa, ##2)\else{}\fi}%
    \def\citeauthoryear##1##2##3{##1 (##3)}\@citedata}
\def\citeA{\def\@citeseppen{-1000}%
    \def\@cite##1##2{(##1\if@tempswa , ##2\fi)}%
    \def\citeauthoryear##1##2##3{##1}\@internalcite}
\def\citeANP{\def\@citeseppen{-1000}%
    \def\@cite##1##2{##1\if@tempswa , ##2\fi}%
    \def\citeauthoryear##1##2##3{##1}\@internalcite}
\def\shortcite{\def\@citeseppen{-1000}%
    \def\@cite##1##2{(##1\if@tempswa , ##2\fi)}%
    \def\citeauthoryear##1##2##3{##2 ##3}\@internalcite}
\def\shortciteNP{\def\@citeseppen{-1000}%
    \def\@cite##1##2{##1\if@tempswa , ##2\fi}%
    \def\citeauthoryear##1##2##3{##2 ##3}\@internalcite}
\def\shortciteN{\def\@citeseppen{-1000}%
    \def\@cite##1##2{##1\if@tempswa, ##2\else{}\fi}%
    \def\citeauthoryear##1##2##3{##2 (##3)}\@citedata}
\def\shortciteA{\def\@citeseppen{-1000}%
    \def\@cite##1##2{(##1\if@tempswa , ##2\fi)}%
    \def\citeauthoryear##1##2##3{##2}\@internalcite}
\def\shortciteANP{\def\@citeseppen{-1000}%
    \def\@cite##1##2{##1\if@tempswa , ##2\fi}%
    \def\citeauthoryear##1##2##3{##2}\@internalcite}
\def\citeyear{\def\@citeseppen{-1000}%
    \def\@cite##1##2{(##1\if@tempswa , ##2\fi)}%
    \def\citeauthoryear##1##2##3{##3}\@citedata}
\def\citeyearNP{\def\@citeseppen{-1000}%
    \def\@cite##1##2{##1\if@tempswa , ##2\fi}%
    \def\citeauthoryear##1##2##3{##3}\@citedata}
%
%
%
\def\@citedata{%
    \@ifnextchar [{\@tempswatrue\@citedatax}%
                  {\@tempswafalse\@citedatax[]}%
}

\def\@citedatax[#1]#2{%
\if@filesw\immediate\write\@auxout{\string\citation{#2}}\fi%
  \def\@citea{}\@cite{\@for\@citeb:=#2\do%
    {\@citea\def\@citea{, }\@ifundefined
       {b@\@citeb}{{\bf ?}%
       \@warning{Citation `\@citeb' on page \thepage \space undefined}}%
{\csname b@\@citeb\endcsname}}}{#1}}%

%
\def\@citex[#1]#2{%
\if@filesw\immediate\write\@auxout{\string\citation{#2}}\fi%
  \def\@citea{}\@cite{\@for\@citeb:=#2\do%
    {\@citea\def\@citea{; }\@ifundefined
       {b@\@citeb}{{\bf ?}%
       \@warning{Citation `\@citeb' on page \thepage \space undefined}}%
{\csname b@\@citeb\endcsname}}}{#1}}%

%
\def\@biblabel#1{}
\makeatother



\newdimen\bibindent
\bibindent=0.0em
\def\thebibliography#1{\section*{\refname}\list
   {}{\settowidth\labelwidth{[#1]}
   \leftmargin\parindent
   \itemindent -\parindent
   \listparindent \itemindent
   \itemsep 0pt
   \parsep 0pt}
   \def\newblock{}
   \sloppy
   \sfcode`\.=1000\relax}

%% file: body.tex
\section{Introduction}

Agent-based modeling (ABM) is a helpful tool in studying epidemic dynamics. \citeN{cajka2010} create a synthetic population and develop an ABM where they assign the agents in the synthetic population to households, workplaces, schools, and other assignments. \shortciteN{ferguson2006strategies} develops a large-scale simulation to study different interventions for a novel epidemic outbreak. \citeN{chao2010flute} develops a simulation for the spread of influenza in a large population, the model is calibrated with historical influenza data. Similarly, \shortciteN{tuite2010MCMC} develops a Markov chain Monte-Carlo simulation model based on historical data of symptoms that are obtained for laboratory-confirmed cases of H1N1 influenza. \citeN{parker2011globalscale} develops a simulation model that can simulate epidemic outbreaks on a global scale (i.e., several billion agents). \shortciteN{grefenstette2013fred} build on these concepts and develop them into a generic framework for epidemic dynamics modeling called FRED (A Framework for Reconstructing Epidemiological Dynamics). \shortciteN{lukens2014fluh1n1} develop an ABM linked with an equation-based, within-host model for influenza in FRED. \shortciteN{potter2012school} develop an ABM to study local school closure policies during an influenza epidemic. \shortciteN{liu2015measle} build a model of Measles transmission using FRED where they study the role of vaccination in preventing Measles outbreaks, this paper has been influential in stressing the importance of vaccination and impacting public policy. Beyond influenza and measles models, \shortciteN{krauland2020cardio} study cardiovascular disease death and its social determinants using the  FRED synthetic population. The latter study demonstrates the possibility of understanding disease risk and consequences of large-scale interventions using FRED synthetic population. 

The computational cost of population-scale, agent-based simulations could be high, and modelers resort to inferential tools to estimate surrogate models that summarize relevant information in the simulations and circumvent having to rely on simulation samples for upstream inference and optimization tasks. Gaussian process regression (GPR) is one such inferential tool. \citeN{fearnhead2014inference} use Gaussian processes in inferring reaction rates of networks which is usually a type of network used in applications of epidemics and biology. \citeN{senanayake2016aaai} and 
\citeN{zimmer2020inficml} develop a GPR model to understand the dynamics of influenza. \citeN{zimmer2017likelihood} use Gaussian processes to calibrate and estimate epidemic parameters and point out advantages over other comparable methods. \citeN{ball2017heterogeneous} considered inference for a SIR epidemic model on a network using GPR based on the covariance function of a branching process. \citeN{buckingham2018gaussian} use GPR for SIR (Susceptible-Infected-Recovered) and SEIR (Susceptible-Exposed-Infected-Recovered) models to estimate underlying parameters. 

We show an application of the Gaussian process regression to FRED-synthesized data from two epidemic models, an influenza model (INF) and an opioid use disorder mode (OUD). Subsequently, we can compare the spatial auto-correlation values for the two diseases and compare their spatial dependencies, keeping everything the same across the two synthetic populations. Both models were constructed over the same geographic location with the same agent's information in the area and starting from the same initial conditions. Thus, we can attribute the observed difference in spatial clustering to the differing disease dynamics, because the remaining conditions are the same for both models using the same synthetic population.

This paper is structured into six sections. In \cref{sec:fred}, we briefly introduce the FRED simulation software and detail how FRED works. In \cref{sec:GPR}, we provide preliminaries about GPR and introduce our notation for GPR with spatial, FRED-synthesized data. In \cref{sec:method}, we will discuss our  modeling assumptions inference and methodology. In \cref{sec:results}, we will discuss the results of our study and its public health implications. Finally, in \cref{sec:conc} we provide concluding remarks and give future work directions.

\section{FRED simulation software}\label{sec:fred}
FRED (Framework for Reconstructing Epidemiological Dynamics) is an agent-based, open-source software that is developed to simulate the temporal and spatial behaviors of epidemics. FRED is developed by the Public Health Dynamics Laboratory (PHDL) at the University of Pittsburgh School of Public Health. FRED was developed originally to study epidemic dynamics, however, it has shown broader potential for population studies that help guide public health policies and interventions. One of the key features of FRED is its synthetic population that is based on the US Census Bureau's public use Microdata files and Census aggregated data which we discuss below \cite{guclu2016agent}.

\subsection{Synthetic Population}
FRED represents every individual in a specified geographic area explicitly. FRED utilizes the US synthetic population database from RTI International \cite{rti}, where the synthetic population contains detailed geographically-allocated categories. The synthetic population has geographically-assigned, synthetic households. Each household has residents, and there are also residents for different facilities (e.g., dorms, prisons, etc.), schools, and mapping of students to schools. Each household, school, and workplace is assigned to a specific geographic region. These assignments reflect an implicit representation of the distance between agents and facilities. Each agent has its own assigned socioeconomic and demographic information (e.g., age, sex, race, employment, etc.) and assigned locations to their activities (e.g., school, workplace, household, etc.) \cite{grefenstette2013fred}.

\subsection{Discrete-time simulation}
At every simulation step, the agent interacts with other agents likely to  share similar daily activities. For example, agents in the same workplace interact with the same colleagues daily. If an infected agent interacts with a susceptible agent, there is a chance for disease transmission from the infected agent to the susceptible one. Additionally, the agent can dynamically change their daily activities (e.g., stop going to work, change locations, etc.). Moreover, FRED allows detailed and effective evaluation of specific interventions as every disease transmission is recorded in the software logs.

\subsection{Agent model}
Each agent has its own attributes (e.g., age, race, gender, employment, school, etc.) and activities' locations (e.g., household, school, workplace, etc.). Newborns are assigned to the same household as their mothers, if children reach school age, they are assigned to a school, and if an agent dies, it is erased from the synthetic population. 

FRED allows agents to take actions especially in contexts related to public health. For example, agents can decide to stay home if sick, prevent their children from attending school if sick, or accept or reject vaccinations. At each step, the agent's actions involve interactions between the agent's internal actions and factors of their environment (e.g., the availability of vaccines given their decision to take it).

\subsection{Disease model}
FRED supports the propagation of one or more infectious diseases in the population. Each disease is determined by specific parameters for transmission, contact, history, etc. For example, typically the agent will pass into three stages for infectious disease, Susceptible, Infected, Recovered, or S-I-R. The agent will move from each state to another based on the transmission rate (defined by the modeler) and also based on each agent's data (e.g., if the agent is registered in a school where the disease is spreading, then the agent will have an increased infection rate).
Epidemics in FRED are propagated by assigning random agents in the population to have the infectious disease as an initialization. It is also possible to modify these assignments based on specified groups (e.g., related to race, geography, age, etc.).

\subsection{Place model}
FRED supposes that all disease-specific interactions between agents occur in a "place", where each type of place specifies a unique environment for spreading infection. The default type of places in FRED include: schools, workplaces, households, etc. Agents are expected to do their daily activities in proximity to assigned locations. However, agents can move to another location during the day. Based on its specific information each agent has specific default daily schedule (e.g., times to go to the workplace, weekends, or annual holiday schedules). Places can also be location-free, allowing us to model social network interactions in FRED.

\section{Gaussian Process Regression}\label{sec:GPR} 
A Gaussian Process (GP) is a type of stochastic process where the value of any random variable within the process follows Gaussian distribution. It can be seen as a generalization of the Gaussian probability distribution (hence the name). Each realization of a Gaussian process is a function $f(\cdot)$, and one way to look at GP is to think of its realizations as long vectors takings values $f(x)$ at specified inputs $x_1,x_2,\ldots$. In this view, a good GP fit contains information about infinite queries and allows us to make inferences about any finite collection of points \cite{williams2006}. Mathematically, a GP can be specified as follows:
\begin{equation}
    f(x) \sim GP(\mu(x),k(x,x'))
\end{equation}
where $\mu(x)$ is the expected value given input location $x$ and $k(x,x')$ is the kernel function of the GP (also known as the covariance function) \cite{duvenaud2014}. Starting from prior parameters and given the observed data, Gaussian process regression (GPR) provides a Bayesian procedure to update GP kernel parameters based on their fit to the observed data. The Bayesian procedure consists of a sequence of iterative steps where a loss function is calculated at every step, and based on the result the model's parameters are updated, until a predefined criterion is met and the procedure stops.

\subsection{Kernels}
A kernel (also known as covariance function, kernel function, or covariance kernel) is a function that defines the similarity between two points $x,x'$ where $x$ and $x'$ are typically vectors in the Euclidean space, but can also be defined for images, graphs, or categorical inputs.

\subsubsection{Kernel types}
There are different types of kernels due to the variety of applications for GP. We will mention kernels that are widely used in general including the one that we use in our model.\\

\noindent{\bf Radial Basis function kernel:} Radial basis function (RBF) kernel (also known as the squared-exponential kernel) is defined as:
\begin{align*}
    k(x,x') = \exp(-\dfrac{d(x,x')^2}{2l^2}),
\end{align*}
where $l$ is the length scale parameter and $d(\cdot,\cdot)$ is the Euclidean distance between the input points. As an advantage, RBF kernel is infinitely differentiable, implying that GP with the corresponding kernel has mean square derivatives and is very smooth. However, RBF has limitations in modeling some applications, especially physical processes where other kernels are recommended.  Nevertheless, RBF kernel is the most widely used kernel in the field \cite{williams2006}.\\

\noindent{\bf Matern Kernel:} The Matern class of kernels is a generalization of RBF kernels.  The kernel is given by:
\begin{equation}
    k(x,x') = \dfrac{1}{\Gamma(\nu)2^{\nu-1}} \left(\dfrac{\sqrt{2\nu}}{l}d(x,x')\right)^\nu K_{\nu}\left(\dfrac{\sqrt{2\nu}}{l}d(x,x')\right),
\end{equation} where $d(\cdot,\cdot)$ is the Euclidean distance between the two points, $K_{\nu}(.)$ is a modified Bessel function and $\Gamma(\cdot)$ is the gamma function. $\nu$ is a parameter that controls the smoothness of the kernel, where smoothness increases  with the value of $\nu$. The kernel becomes especially simple when $\nu$ is a half-integer: $\nu=p+1/2$, where $p$ is a non-negative integer. For example when $\nu = 3/2$ the kernel becomes:
\begin{equation}
    k_{\nu=3/2}(x,x') = \left(1+\dfrac{\sqrt{3}*d(x,x')}{l}\right)\exp\left(-\dfrac{\sqrt{3}*d(x,x')}{l}\right).
\end{equation}
Moreover, as $\nu \rightarrow \infty$ the Matern kernel converges to the RBF kernel. 

\subsubsection{Combining kernels}
In some applications a single kernel may not be expressive enough and custom kernel functions are developed. Usually developing this type of custom kernel is by combining already known kernels to get the desired properties, for example, by adding two kernels as follows:
\begin{align*}
    k_a+k_b= k_a(x,x')+k_b(x,x'),
\end{align*}
or by multiplying them together:
\begin{align*}
    k_a*k_b= k_a(x,x')\cdot k_b(x,x').
\end{align*}
It is worth mentioning that the above examples are in one dimension, but creating custom kernels can be done also over multi-dimensional inputs by adding or multiplying different kernels over different dimensions \cite{duvenaud2014}.

\subsection{Spatial auto-correlation}
Spatial auto-correlation measures the degree of correlation between the observed values for spatial units. Moran's I is a common measure for evaluating the level of spatial auto-correlation \cite{moran1950test}. Given a set of features, Moran's I statistic evaluates whether the observed pattern shows significant clustering or if it can be adequately described as random, with values close to 1 or -1 indicating that the location data is clustered positively (occurring more closely to each other than would be expected from a random spread) or negatively (spread out more than expected from a random distribution of locations). A Moran's I value around zero indicates that there is no significant evidence for clustering and the spatial spread of events' locations can be adequately described as random. Moran's I is given by:
\begin{equation}
    I= \frac{N\sum^{N}_{i=1}\sum^{N}_{j=1}w_{ij}(x_i-\bar{x})(x_j-\bar{x})}{W\sum^{N}_{i=1} (x_i-\bar{x})^2},
    \label{eq:moran}
\end{equation} where $N$ is the number of spatial units indexed by $i$ and $j$, $x$ is the variable of the pattern (e.g., disease incidence or prevalence at different locations), $\bar{x} = (\sum_{i=1}^{N}x_i)/N$ is the mean of the variable $x$, $w_{ij}$ represents the spatial weight between $i$ and $j$, and $W = \sum_{i=1}^{N}\sum_{j=1}^{N}w_{ij}$ is the summation of all  $w_{ij}$ values that constitute the spatial weight matrix defined next.\\

\noindent{\bf Spatial weight matrix:} The spatial weight matrix, $(W_{ij})_{N\times N}$, is a representation of the spatial structure over the data. Binary and variable weighting are two main ways of specifying the spatial weights between locations. Examples of binary weights include: contiguity-based, fixed distance, $K$ nearest neighbours, etc. For variable weighting inverse distance is one of the main techniques.
In binary weighting, contiguity means that two spatial units share a common boundary (with von Neumann and Moore being the two main ways for determining boundaries). Moore defines neighbours by the existing edge between the two spatial units, which makes it usually defines four neighbours while von Neumann defines neighbours by the existing edge or vertices between the spatial units, which usually ends up with more neighbours compared to the Moore method. In this paper, the calculation of the weights is based on the distance between ZIP code centers.


\section{Methodology}\label{sec:method}
\subsection{Simulated models}
In this section we will discuss the two simulation models used in our study, both models were built using FRED. The first model is a little simpler than the second one, however, it will help illustrate our points.
\subsubsection{Influenza model}
\begin{figure}[h]
    \centering
    \includegraphics[scale=0.4]{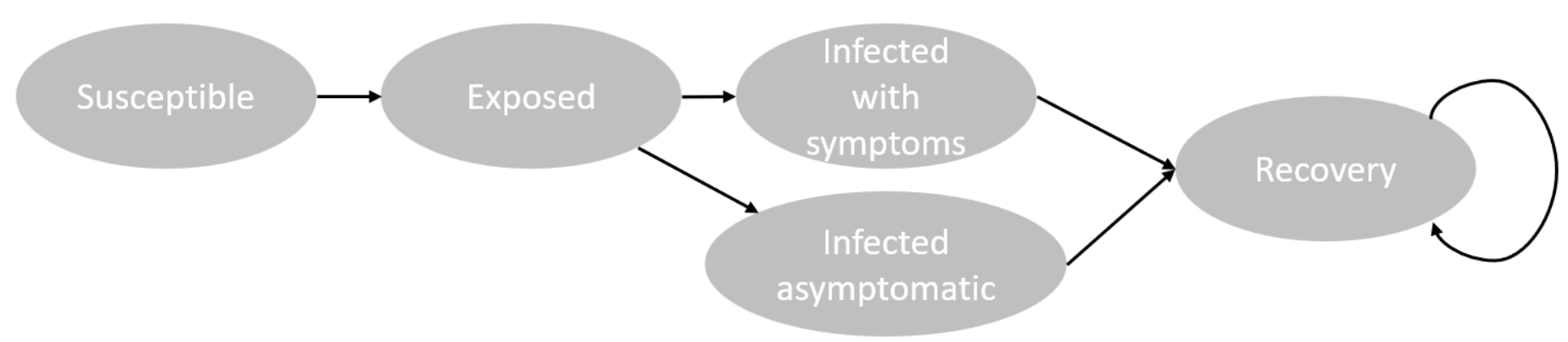}
    \caption{State transition diagram for the INF model.}
    \label{fig:infFlow}
\end{figure}

Influenza (INF) model is one of the well-studied models provided with the installation of FRED simulation. INF model has five defined states as shown in \cref{fig:infFlow}: Susceptible, Exposed, Infected with Symptoms, Infected Asymptomatic, and Recovery. The transition of the agent from state to state is based on explicitly defined probabilities, moreover, the agent's residing time in each state is explicitly defined for each state. We simulated the INF model over Allegheny County, Pennsylvania. The INF model can be found on FRED library on the following link: https://github.com/PublicHealthDynamicsLab/FRED/tree/FRED-v5.0.0/library/Influenza. 

\subsubsection{Opioid Use Disorder model}
The Opioid Use Disorder (OUD) model is a model developed to understand the OUD epidemic across the U.S. The growth of substance use disorder and drug overdose deaths is a growing public health concern in the U.S. Currently. Opioids are the leading cause of drug overdose deaths in the U.S. \citeN{CDCOpioids} (including prescription opioids, heroin, and synthetic opioids). \citeN{jalal2018opioiddynamics} studies the epidemic dynamics over the last 40 years and concluded that the current opioid overdose deaths wave is part of a long-trend that is sustained over several decades stressing the importance of studying the epidemic dynamics. The OUD model that we use in this paper was developed in the Public Health Dynamics Laboratory at the University of Pittsburgh, based on data provided by Centers for Disease Control and Prevention (CDC) as a part of their sponsored research project. Similar to the INF model, OUD model has explicitly defined transition probabilities between different states and dwell times for each agent at different states. Nevertheless, OUD model differs from INF in that its simulation update steps occur every month. Similar to the INF model and to have standard comparisons, the OUD model was simulated over the synthetic population of Allegheny County, PA. Both simulations were conducted over the same time frame between Jan 1st, 2016 to Dec 31st, 2017. The reason for this specific time frame is that OUD model state transitions were calibrated to the real data over the same period.
\begin{figure}[h]
    \centering
    \includegraphics[scale=0.45]{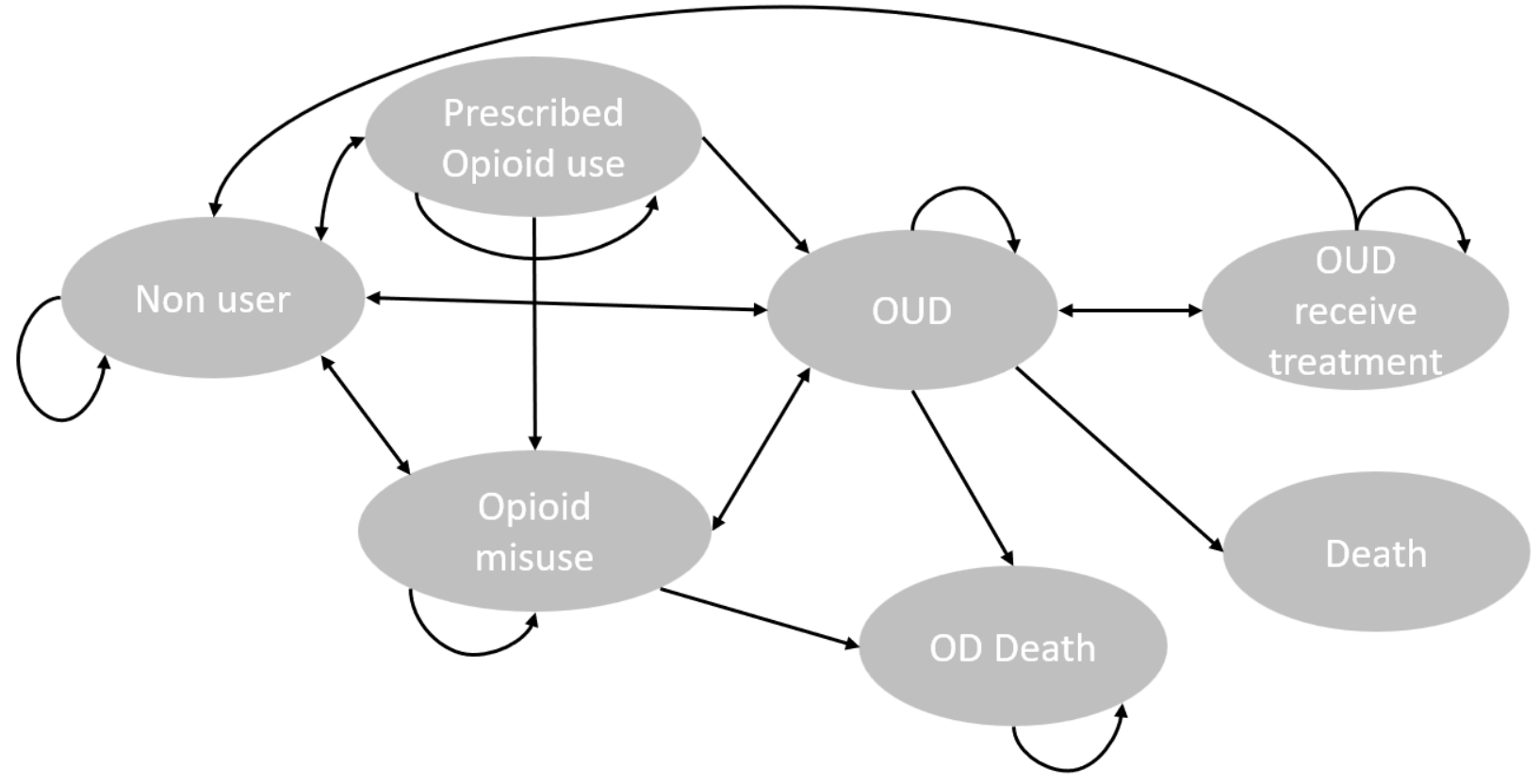}
    \caption{State transition diagram for the OUD model.}
    \label{fig:oudFlow}
\end{figure}

\subsection{FRED Simulation Results}
The INF and OUD models resulted in detailed records for each specified state over the simulated time window. We picked a single state from each simulation model to compare their behaviors. In the INF model, we picked the Infected with Symptoms (IS) state numbers over the simulated time window, while for the OUD model, we picked the Overdose Deaths (ODD) state over the same time window. Therefore, the behaviors we were comparing are the number of Infected with Symptoms numbers monthly from INF model against the number of ODD monthly from the OUD model. We aimed to use an inferential technique to measure statistical differences between the spatial spread of the epidemic dynamics in each case (i.e., spatial spread of the number of IS in the INF model and the number of ODD in the OUD model). In this study, we used GP as an inference model for our target parameters, where we select a single month (simulation step) to study the spatial spread. We avoided the starting and ending months in the two-year simulation period. For choosing a kernel, we tested different kernels to better infer the target parameters. To decide which kernel to use we tested selected kernels on a separate set (i.e., test-set) and selected the kernel that has the lowest means square error (MSE) for each model, \cref{tab:compare} shows MSE for different kernels RBF, Matern, and a combined kernel for the INF and OUD models.
The training was done using Python language (https://github.com/abdulrahmanfci/gpr-abm) and the GPFlow library \citeN{GPflow2017} for GP regression. 

\begin{table}[h]
\vspace{5pt}
    \centering
    \begin{tabular}{|c|c|c|c|}
    \hline
        Model&  RBF & Matern & RBF*Matern\\ \hline
      INF&  \textbf{48.2} & 51.06 & 51.9 \\ \hline
      OUD& 0.77  & \textbf{0.59} & 0.592 \\
      \hline
    \end{tabular}
    \caption{Showing the MSE for different kernels tested for the two models (INF and OUD).}
    \label{tab:compare}
\end{table}

\section{Results and Discussion}\label{sec:results}
\begin{figure}[h]
    \centering
    \subfloat[\centering simulated data map]{{\includegraphics[width=7.5cm]{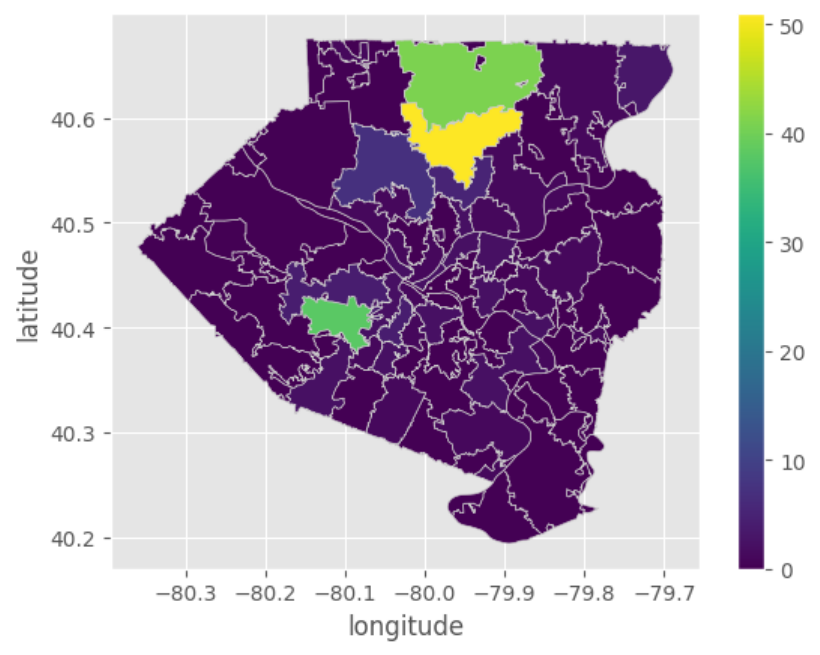} }}%
    \qquad
    \subfloat[\centering GP regression for the map]{{\includegraphics[width=7.5cm]{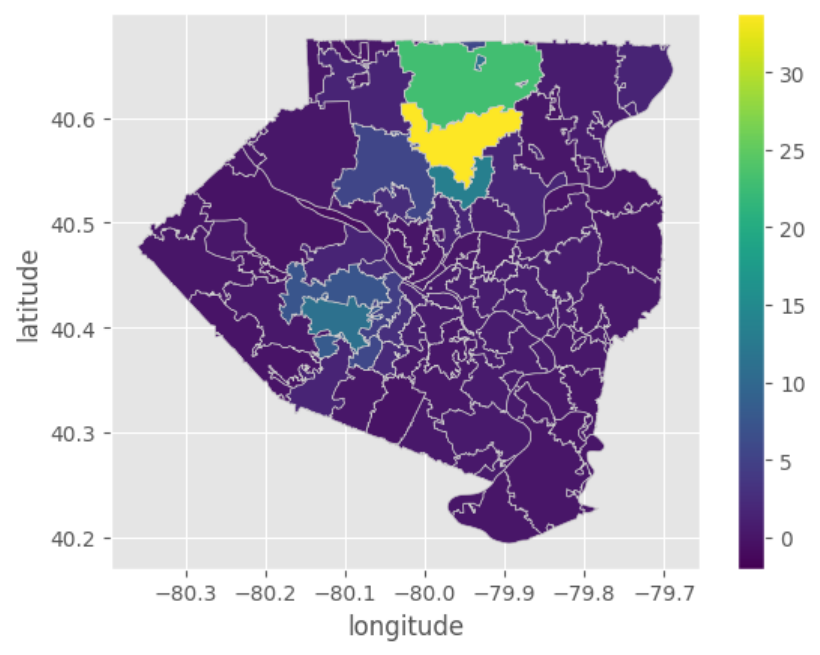} }}%
    \caption{INF model simulated data versus GP regression for Allegheny county, PA.}%
    \label{fig:infmodel}%
\end{figure}

\begin{figure}[h]
    \centering
    \subfloat[\centering INF model sample]{{\includegraphics[width=7.5cm]{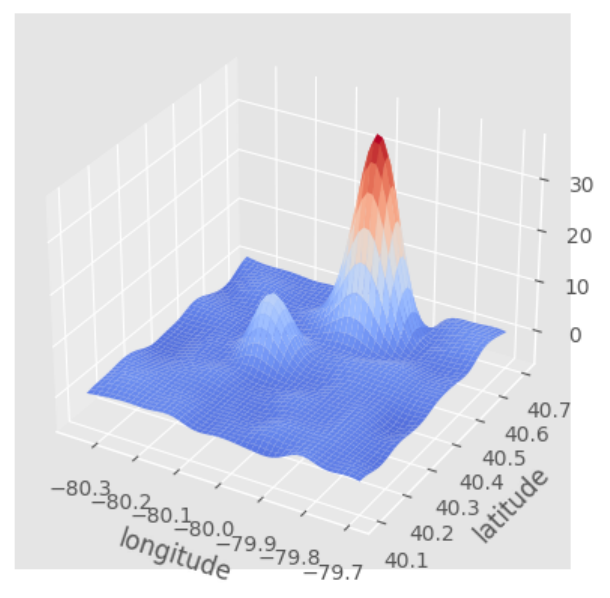} }}%
    \qquad
    \subfloat[\centering OUD model sample]{{\includegraphics[width=7.5cm]{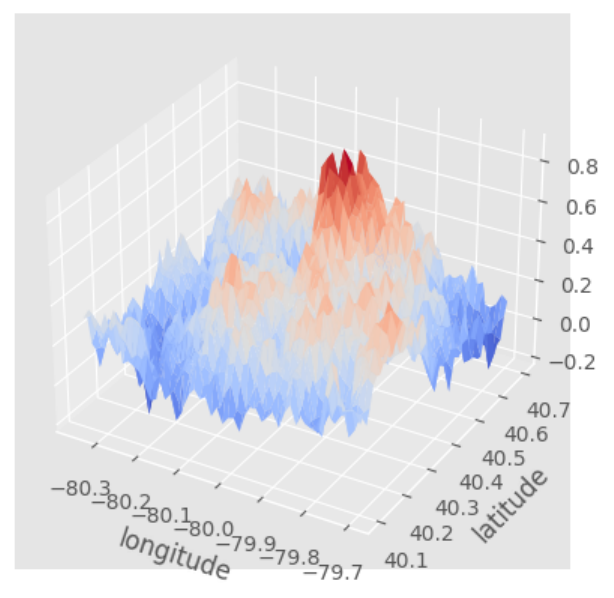} }}%
    \caption{Posterior samples from GPR for different values of x and y in INF and OUD models respectively}%
    \label{fig:sample}%
\end{figure}

The FRED-synthesized data are the list of coordinates where the IS and ODD happened at each simulation step. To better visualize the data we used the map of Allegheny county, PA ZIP codes to assign each incident location to its ZIP code. For example, if the first month of the INF model had two IS incidents happening at coordinates $(x_1,y_1)$ and $(x_2,y_2)$ within the 15261 ZIP code, then we assign a count of two to that ZIP code area. Afterward, we fitted a GP to the processed location data from the FRED output, taking the input, $x$, as Allegheny County's, PA ZIP code areas and the response $y$ as the number of events in each ZIP code area. \cref{fig:infmodel} shows the simulated data for the INF model over the Allegheny County, PA ZIP codes map with the corresponding GP fitted values in part (b). The values shown in (b) are the mean of the posterior distribution. Because the INF model has relatively abundant numbers of IS over the region, a per-made RBF kernel was capable of modeling the INF disease's behavior. \cref{fig:sample} (a) shows a sample from the posterior distribution plotted over the longitude and latitude coordinates.

\begin{figure}[h]
    \centering
    \subfloat[\centering simulated data map]{{\includegraphics[width=7.5cm]{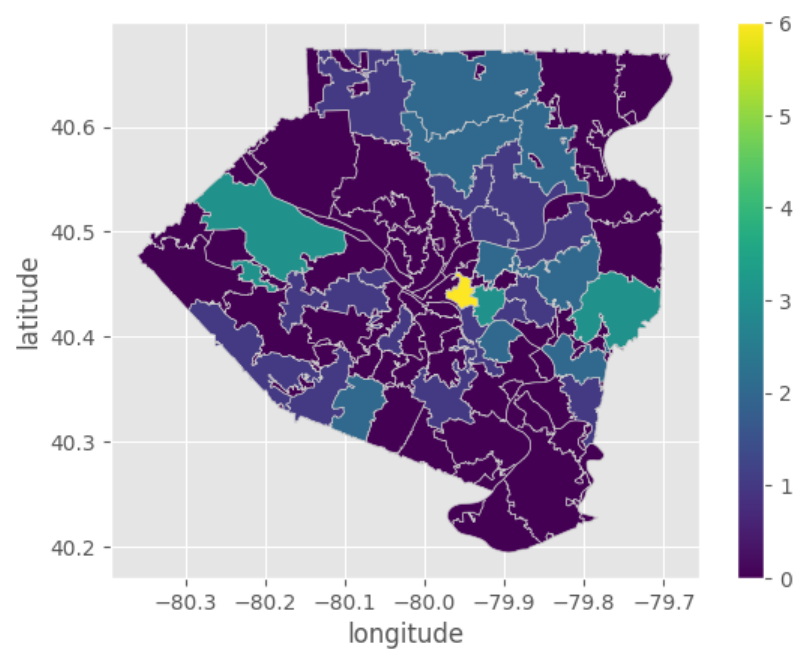} }}%
    \qquad
    \subfloat[\centering GP regression for the map]{{\includegraphics[width=7.5cm]{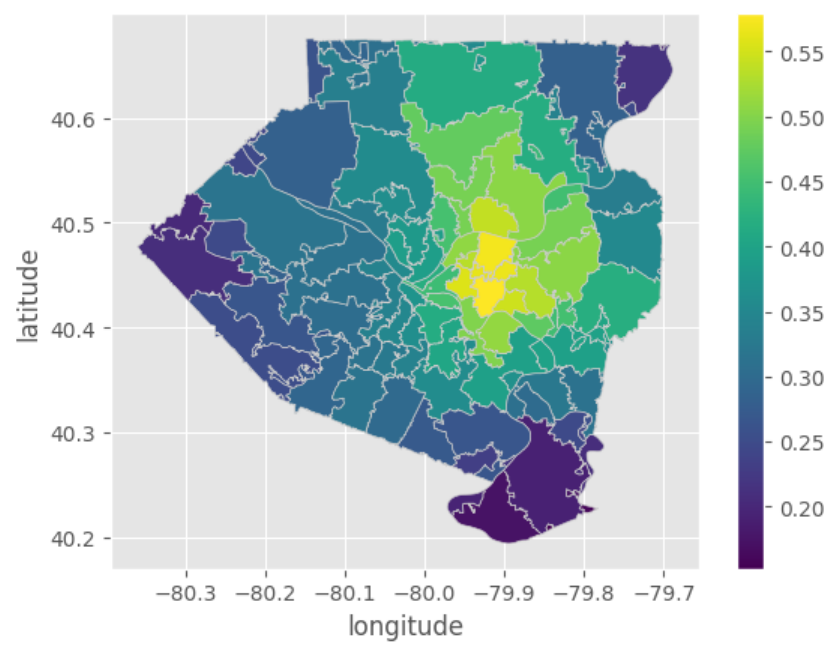} }}%
    \caption{OUD model simulated data and GP regression for Allegheny county, PA.}%
    \label{fig:oudmodel}%
\end{figure}

Similarly for the OUD model the FRED-synthesized data are the list of coordinates representing death events at each simulation step. We applied the same post-processing to map OD death locations to their ZIP code areas. In the case of the OUD model, we tested different GP kernels and combinations to improve the GP fit. The Matern kernel provided the best fit to the simulated data. In contrast to the INF model, the OUD model incident data are sparser and more spread out over the region. \cref{fig:oudmodel} shows the simulated data for the OUD model for Allegheny County, PA and the corresponding GP fitted values in part (b), where the values shown in (b) are the mean of the GP posterior distribution. We can see from \cref{fig:oudmodel} that the fitted values from GPR captured the center of the spread, however, they failed to pin down the location of the surge. This could be due to the sparsity of the ODD incident locations which makes tuning the parameters of the GP for the OUD model harder than for the INF model. Similarly, \cref{fig:sample} (b) shows a sample from the posterior distribution plotted over the longitude and latitude coordinates.

\begin{figure}[h]
    \centering
    \includegraphics[scale=0.5]{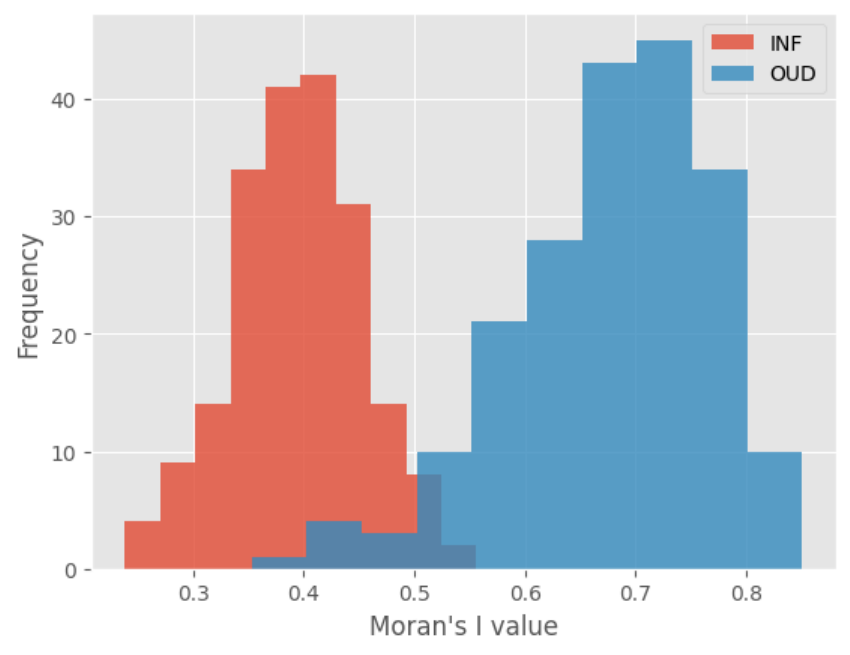}
    \caption{Moran's I test values for INF and OUD.}
    \label{fig:moran}
\end{figure}

\vspace{5pt}
\noindent{\bf Spatial Autocorrelation Analysis:} To demonstrate the utility of GPR for inferring properties of the underlying epidemic dynamics, we conducted a spatial auto-correlation analysis using our GPR results. We used Moran's I statistics (\cref{eq:moran}) to compare our two fitted GPs. The mean values for Moran's I statistics over 200 random draws for INF was 0.4 and for OUD was 0.68. We conducted a difference in means test where we assumed Gaussian distribution for population means. This resulted in a $95\%$ confidence interval of $\mu_{INF}-\mu_{OUD} \in \left[-0.30,-0.27\right]$ for the difference in means. \cref{fig:moran} shows the distribution of Moran's I statistics for each model. The results indicate that incident locations in the INF model are less clustered than the OUD model. This could be verified by inspecting \cref{fig:infmodel,fig:oudmodel} where occurrences of ZIP codes with high incident rates are fewer in the former case. Moreover, the results for OUD indicate more spatial correlation between adjacent ZIP codes compared to the INF model which does not show a strong relationship between incident counts in neighbouring ZIP codes. The significant difference between spatial clustering may be attributed to strong socioeconomic correlates that derive the OUD spread, whereas INF is primarily driven by the spatial spread of the viral disease agents. 

\noindent This result shows the potential of GPR to be used in inferring epidemic dynamics. Such capabilities are especially useful when simulating many interventions across large populations where GPR can be used to interpolate across population characteristics and treatment conditions, thus circumventing the need to simulate each condition across the entire population \cite{ahmed}. 
In this part, we used functions from PySAL library \citeN{pysal2007} --- PySAL is a library specialized in providing geospatial tools and applications for spatial analysis in Python.

\section{Conclusion}\label{sec:conc}
Because the INF and OUD models were tested on the same synthetic population over the same time window and geographic location, the observed differences between the fitted parameters of GP, and in particular, the different spatial auto-correlation measures can be attributed completely to the differing epidemic dynamics --- all the relevant conditions in the synthetic population (whom each agent lives with within their household, whom they meet in their workplaces, etc.) are otherwise fixed. Such sharp inferences are typically impossible with real data due to the variety of factors that confound our comparisons. In this paper, we showed that the same area of Allegheny County, PA, with the same agent's information will end up having different inferences for each disease due to their differing dynamics. These results demonstrate the potential of FRED simulations for comparing treatment conditions on identical synthetic populations, and thus controlling for every confounding factor that may otherwise compromise our contrasts. We also illustrated the utility of Gaussian process regression as an inference tool for comparing epidemic dynamics on FRED-synthesized data. Such surrogate modeling techniques help us circumvent the computational burden of generating extensive simulation samples from large population dynamics. 

\noindent This initial analysis opens several directions for future work, including a comparison of more complex spatial dependencies, e.g., spillovers across adjacent counties that can directly inform public health policy and intervention design \shortciteN{holtz2020interdependence}, and extending the methodology for spatiotemporal regression. Our current analysis focused only on spatial relations and investigating inferences with spatiotemporal GPR is an important future direction.

\section*{Data and code}
In this study, we are not able to share detailed data about the OUD model for contractual reasons. The repository link for the paper's code can be found at https://github.com/abdulrahmanfci/gpr-abm. 

\section*{ACKNOWLEDGMENTS}
This research was funded by contract 75D30121C12574 from the Centers for Disease Control and Prevention. The findings and conclusions in this work are those of the authors and do not necessarily represent the official position of the Centers for Disease Control and Prevention.

This research was supported in part by the University of Pittsburgh Center for Research Computing, RRID:SCR\textunderscore022735, through the resources provided. Specifically, this work used the HTC and VIZ clusters, which are supported by NIH award number S10OD028483.


\footnotesize

\bibliographystyle{wsc}

\bibliography{demobib}

\section*{AUTHOR BIOGRAPHIES}


\noindent {\bf Abdulrahman A. Ahmed} is a PhD student in the Department of Industrial Engineering at University of Pittsburgh. He obtained his MSc and BSc in Operations Research and Computer Science from Cairo University. His current research is on using surrogate models to develop inference procedures on complex sociotechnical systems with a focus on public health. His email address is \email{aba173@pitt.edu}. \\

\noindent {\bf M. Amin Rahimian} is an Assistant Professor in the Department of Industrial Engineering at University of Pittsburgh. His current research focus is on challenges of inference and intervention design in complex, large-scale sociotechnical systems with applications ranging from social networks and e-commerce, to public health. His email address is \email{rahimian@pitt.edu}, and his website is \url{https://aminrahimian.github.io}.\\

\noindent {\bf Mark S. Roberts} is a distinguished professor in the Department of Health Policy and Management at University of Pittsburgh, and holds secondary appointments in Medicine, Industrial Engineering, Business Administration, and Clinical and Translational Science. His recent research has concentrated in the use of mathematical methods from operations research and management science, including Markov Decision Processes, Discrete Event, and Agent-based Simulation. His email address is \email{mroberts@pitt.edu} and his page is: \url{https://www.sph.pitt.edu/directory/mark-roberts}.\\
